\documentclass[prl,superscriptaddress]{revtex4-2}
\pdfoutput=1
\usepackage{times}
\usepackage{graphicx}
\usepackage{float}
\usepackage{amssymb}
\usepackage{mathrsfs}
\usepackage{hyperref}
\usepackage{amsmath, amssymb,textcomp,calc,capt-of,ifthen}
\usepackage[english]{babel}
\usepackage{xcolor,ulem}
\hypersetup{
           breaklinks=true,   
           colorlinks=true,   
           pdfusetitle=true,  
}

\graphicspath{{figure/}}

\begin{document}

\title{Entanglement purification and protection in a superconducting quantum network}

\author{Haoxiong Yan}
\thanks{These two authors contributed equally}
\affiliation{Pritzker School of Molecular Engineering, University of Chicago, Chicago IL 60637, USA}
\author{Youpeng Zhong}
\thanks{These two authors contributed equally}
\altaffiliation[Present Address: ]{Shenzhen Institute for Quantum Science and Engineering, Southern University of Science and Technology, Shenzhen 518055, China}
\affiliation{Pritzker School of Molecular Engineering, University of Chicago, Chicago IL 60637, USA}
\author{Hung-Shen Chang}
\affiliation{Pritzker School of Molecular Engineering, University of Chicago, Chicago IL 60637, USA}
\author{Audrey Bienfait}
\altaffiliation[Present Address: ]{Universit\'{e} de Lyon, ENS de Lyon, Universit\'{e} Claude Bernard, CNRS, Laboratoire de Physique, F-69342 Lyon, France}
\affiliation{Pritzker School of Molecular Engineering, University of Chicago, Chicago IL 60637, USA}
\author{Ming-Han Chou}
\affiliation{Pritzker School of Molecular Engineering, University of Chicago, Chicago IL 60637, USA}
\affiliation{Department of Physics, University of Chicago, Chicago IL 60637, USA}
\author{Christopher R. Conner}
\affiliation{Pritzker School of Molecular Engineering, University of Chicago, Chicago IL 60637, USA}
\author{\'Etienne Dumur}
\altaffiliation[Present Address: ]{Universit\'{e} Grenoble Alpes, CEA, INAC-Pheliqs, 38000 Grenoble, France}
\affiliation{Pritzker School of Molecular Engineering, University of Chicago, Chicago IL 60637, USA}
\affiliation{Center for Molecular Engineering and Material Science Division, Argonne National Laboratory, Lemont IL 60439, USA}
\author{Joel Grebel}
\affiliation{Pritzker School of Molecular Engineering, University of Chicago, Chicago IL 60637, USA}
\author{Rhys G. Povey}
\affiliation{Pritzker School of Molecular Engineering, University of Chicago, Chicago IL 60637, USA}
\affiliation{Department of Physics, University of Chicago, Chicago IL 60637, USA}
\author{Andrew N. Cleland}
\email{anc@uchicago.edu}
\affiliation{Pritzker School of Molecular Engineering, University of Chicago, Chicago IL 60637, USA}
\affiliation{Center for Molecular Engineering and Material Science Division, Argonne National Laboratory, Lemont IL 60439, USA}

\date{\today}


\begin{abstract}
High-fidelity quantum entanglement is a key resource for quantum communication and distributed quantum computing, enabling quantum state teleportation, dense coding, and quantum encryption. Any sources of decoherence in the communication channel however degrade entanglement fidelity, thereby increasing the error rates of entangled state protocols. Entanglement purification provides a method to alleviate these non-idealities, by distilling impure states into higher-fidelity entangled states. Here we demonstrate the entanglement purification of Bell pairs shared between two remote superconducting quantum nodes connected by a moderately lossy, 1-meter long superconducting communication cable. We use a purification process to correct the dominant amplitude damping errors caused by transmission through the cable, with fractional increases in fidelity as large as $25\%$, achieved for higher damping errors. The best final fidelity the purification achieves is $94.09\pm 0.98\%$. In addition, we use both dynamical decoupling and Rabi driving to protect the entangled states from local noise, increasing the effective qubit dephasing time by a factor of four, from $3~\rm \mu s$ to $12~\rm\mu s$. These methods demonstrate the potential for the generation and preservation of very high-fidelity entanglement in a superconducting quantum communication network.
\end{abstract}
\maketitle

Superconducting qubits are a favored hardware platform for implementing quantum computation, with extant demonstrations of circuits with up to $\sim 10^2$ physical qubits \cite{Barends2014,Arute2019,Jurcevic2021,Gong2021,Wu2021}. However, there remain significant practical challenges in scaling up to the much larger qubit numbers needed for error correction and for the implementation of useful algorithms \cite{Fowler2012, Hertzberg2021}. Distributed quantum computing provides one path to scaling up, by connecting large numbers of small-scale quantum processors in a quantum network \cite{Gottesman1999, Jiang2007, Kimble2008, Monroe2014}. Initial steps have been taken to link small superconducting processors using superconducting transmission lines \cite{Kurpiers2018, Axline2018, CampagneIbarcq2018, Leung2019, Zhong2019, Burkhart2021, Magnard2020,Zhong2021,Gold2021}, as well as efforts to build coherent microwave-to-optical transducers for optical communication \cite{han2020, Mirhosseini2020a}. However, other than monolithic demonstrations on a single chip \cite{Zhong2019}, these have not realized the high fidelity entanglement needed for quantum information applications, with photon loss dominating the degradation of coherence during transmission through the cable interconnects, and decay due to local noise limiting the quantum state storage time.

Here we describe a superconducting quantum network with two physically-separated nodes, each including three superconducting qubits, connected by a 1-meter superconducting coaxial cable \cite{Zhong2021}. Using this setup, we can deterministically generate high-fidelity Bell pairs shared between the two nodes by sending microwave photons through the coaxial cable. Amplitude damping of the microwave photons however limits the fidelity of the entangled pairs. Here we demonstrate the use of an entanglement purification protocol \cite{Bennett1996} to correct these errors. Entanglement purification via distillation has been demonstrated in linear optics \cite{Pan2001, Pan2003, Hu2021,Ecker2021}, as well as with trapped ions \cite{Reichle2006} and defects in diamond \cite{Kalb2017}. In superconducting qubits, mitigating the photon loss in a communication channel has been achieved using adiabatic methods \cite{Chang2020} as well as through error-correctable qubits \cite{Burkhart2021}. In contrast to adiabatic protocols which require remote synchronization \cite{Vitanov2017, Chang2020}, purification protocols can achieve near unit-fidelity Bell states using only local operations. The purification performance should be similar to protocols using error-correctable qubits, but with no additional requirements for intricate qubit encoding and quantum non-demolition measurements \cite{Duer2007, Burkhart2021}. Here we show that amplitude damping errors can be effectively corrected by a purification protocol including measurement and post-selection. In addition, we use dynamical decoupling (DD) and Rabi driving (RD) \cite{Carr1954, Meiboom1958} to protect the entangled states from local decoherence \cite{Bylander2011,Pokharel2018}, effectively increasing the qubit $T_2$ lifetime by a factor of four, from $3~\rm \mu s$ to $12~\rm\mu s$. These results provide one possible route for the implementation of high-fidelity distributed quantum computing \cite{Gottesman1999, Monroe2014, Fitzsimons2017, Chou2018, Wan2019}.

\begin{figure}
    \centering
     \includegraphics[width=8.6cm]{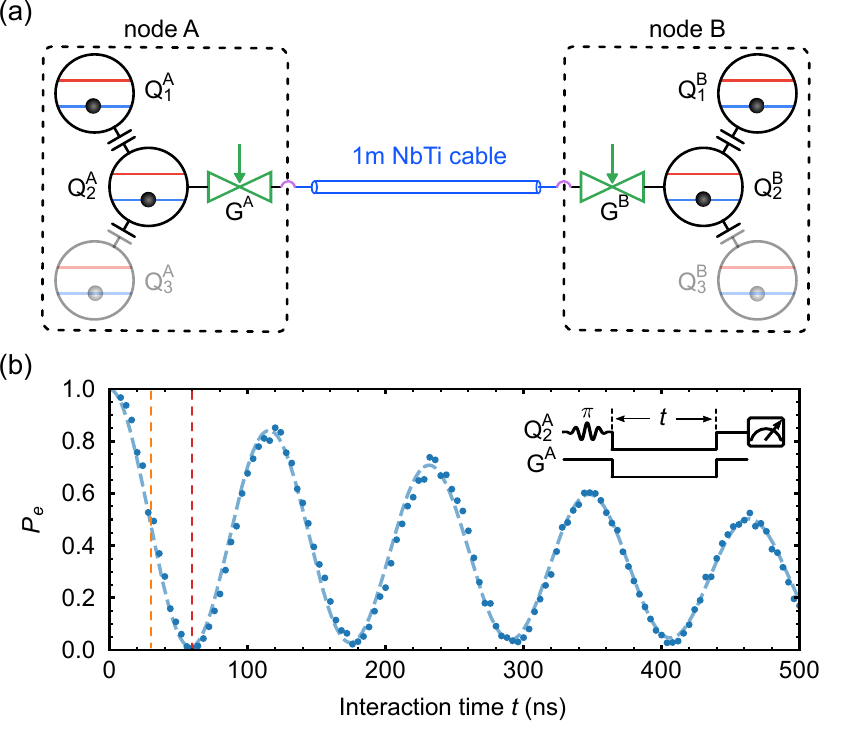}
     \caption{\label{fig:1}Device design and vacuum Rabi oscillations. (a) Schematic of the quantum network, comprising two nodes $A$ and $B$, each with three capacitively-coupled xmon qubits $Q_i^{A,B}$ ($i= 1, 2, 3$). The center qubit $Q_2^{A,B}$ in each node is connected to a 1-meter long superconducting NbTi coaxial cable through a tunable coupler $G^{A,B}$. Qubits $Q_3^{A,B}$ are not used in this experiment. (b) Vacuum Rabi oscillations between $Q_2^A$ and the 5.806 GHz communication mode $C$, measured with coupling strength set to $g^A/2\pi=4.3~\rm MHz$. Inset shows pulse sequence, where after exciting $Q_2^A$, the qubit is tuned into resonance with the communication mode while simultaneously turning on the coupler $G^A$. Blue dashed line is from numerical simulations. The orange and red dashed lines indicate the times for completing a half-swap and a full swap of the excitation in $Q_2^A$ to the communication mode.}
\end{figure}

An overview of the experiment is shown in Fig.~\ref{fig:1}, with a schematic in Fig. \ref{fig:1}(a), previously described in Ref. \cite{Zhong2021}. The system comprises two quantum network nodes $A$ and $B$, where each node includes three capacitively-coupled superconducting qubits $Q_i^k$ ($i=1,2,3$; $k=A,B$), based on the xmon design \cite{Koch2007, Barends2013}. The central qubit $Q_2^k$ in each node is directly coupled to a 1 m-long niobium-titanium (NbTi) superconducting coaxial cable via a tunable coupler $G^k$ \cite{Chen2014}. The tunable coupler is connected by superconducting aluminum wirebonds to the center and ground of the coaxial cable \cite{Zhong2021}, effectively forming a Fabry-P\'erot cavity with the tunable couplers serving as variable mirrors at either end. When the tunable couplers are turned to near zero coupling strength, the Fabry-P\'erot modes have a free spectral range $\omega_{\rm FSR}/2 \pi = 105$ MHz. Here we use the mode at $5.806$~GHz mode for communication, with an energy lifetime of $T_{1r}=477~\rm ns$. The qubits each have a relaxation time of $T_1\approx 10~\rm\mu s$ and dephasing time $T_{2}\approx 3~\mu s$. Vacuum Rabi oscillations between $Q_2^A$ and the communication mode $C$ are shown in Fig.~\ref{fig:1}(b), where we excite $Q_2^A$, then turn on the coupler to coupling strength $g_A/2\pi=4.3~\rm MHz$ while tuning the qubit into resonance with the communication mode; the other coupler is left off, effectively acting as a high-reflectance mirror. The qubit's excited state probability $P_e$ is shown as a function of the interaction time $t$. More details can be found in Refs. \cite{Zhong2021,SupplementaryMaterial}.

\begin{figure*}
    \centering
    \includegraphics[width=12.9cm]{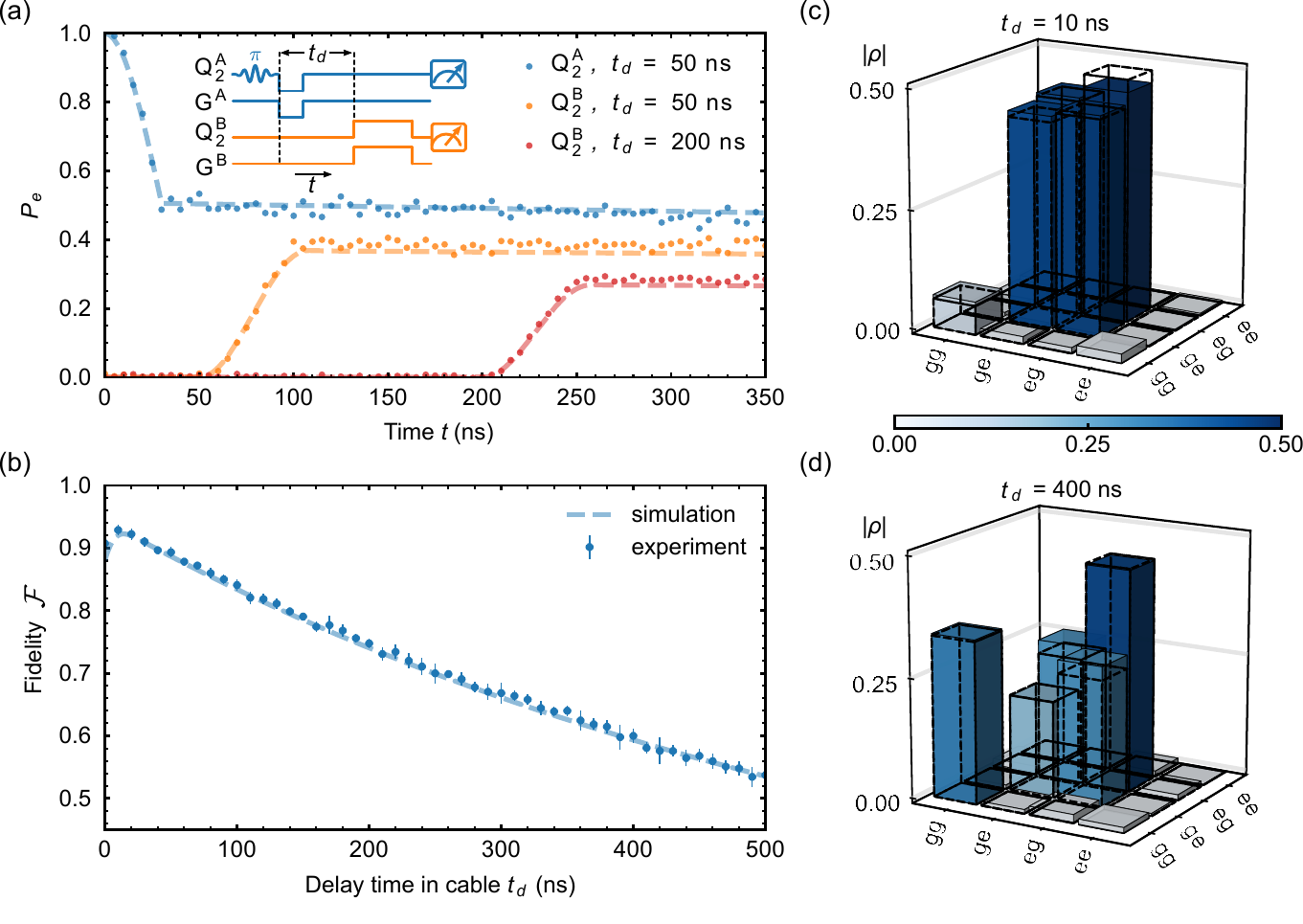}
    \caption{\label{fig:2}Deterministic Bell state generation. (a) Inset: Pulse sequence for Bell state generation, including the delay time $t_{\rm d}$ that the excitation resides in the communication mode $C$. Main plot shows $|e\rangle$ state population $P_e$ in $Q_2^A$ (blue), and in $Q_2^B$ for different delay times in the cable $t_{\mathrm{d}} = 50~\mathrm{ns}$ (orange) and $t_{\rm d} = 200~\mathrm{ns}$ (red). (b) Bell state fidelity $\mathcal F$ as a function of delay time $t_{\rm d}$. (c) Bell state tomography for $t_{\mathrm{d}}=10~\mathrm{ns}$ and (d) $t_{\mathrm{d}}=400~\mathrm{ns}$. Blue dashed lines in (b) and (b), and dashed outline boxes in (c) and (d), are results from numerical simulations.}
\end{figure*}

In Fig.~\ref{fig:2} we display the deterministic generation of a Bell state distributed between nodes $A$ and $B$. Using the tunable coupler, we swap a ``half-photon'' from $Q_2^A$ to $Q_2^B$ with the communication mode $C$ as an intermediate bus. The pulse sequence is shown inset in Fig. \ref{fig:2}(a), where we first apply a $\pi$ pulse to bring $Q_2^A$ from its ground state $|g\rangle$ to its first excited state $|e\rangle$, then turn on $Q_2^A$'s coupler $G^A$ to coupling strength $g^A/2\pi = 4.3~\rm MHz$ while tuning $Q_2^A$ into resonance with the communication mode $C$. The swap time for a full photon emission ($|e0g\rangle \to i|g1g\rangle$, representing states as $|Q_2^A C Q_2^B\rangle$) is $\sim 60~\rm ns$. Here we turn on the coupling for $30~\rm ns$, which swaps a half-excitation to the cable mode (ideally, $|e0g\rangle \to (|e0g\rangle + i|g1g\rangle)/\sqrt{2}$). We then turn off the coupler $G^A$, and after a time delay $t_{\rm d}$, set $G^B$'s coupling strength to $g^B/2\pi = 4.3~\rm MHz$ while tuning $Q_2^B$ into resonance with the communication mode. After a $60~\rm ns$ full swap, this generates a Bell state between $Q_2^A$ and $Q_2^B$, ideally $|\psi^-\rangle = (|eg\rangle - |ge\rangle)/\sqrt{2}$ (writing the two-qubit state as $|Q_2^A Q_2^B\rangle$). In Fig.~\ref{fig:2}(a) we show the excited state probability for $Q_2^B$ for two different delay times, $t_{\rm d} = 50$~ns (orange) and 200~ns (red), along with $Q_2^A$ (blue). These data clearly show the reduction in $Q_2^B$'s excited state probability $P_e$ with delay time $t_{\rm d}$.

In Fig.~\ref{fig:2}(b) we display the effect of the delay time in the cable $t_{\rm d}$ on the Bell state fidelity ${\mathcal F}$, defined as $\mathcal{F} = \langle\psi^-|\rho|\psi^-\rangle$, displaying the two-qubit density matrix $\rho$ measured using state tomography \cite{Steffen2006}. Numerical simulations (dashed blue line; see \cite{SupplementaryMaterial}) are in good agreement with the measurements. In Fig.~\ref{fig:2}(c-d) we show the measured density matrices for the time delays $t_{\rm d}=10~\rm ns$ and $t_{\mathrm{d}}=400~\rm ns$. For the data in Fig.~\ref{fig:2}(c), the measured fidelity to the ideal Bell state is $\mathcal{F} = 92.89\pm 0.85\%$, close to the numerical simulation result $\mathcal{F}^{\rm sim} = 92.01\%$. The data indicate that the dominant infidelity is due to damping errors ($|1\rangle \to |0\rangle$) in the cable, which increase with delay time in the cable, with a much smaller contribution from phase errors in the qubits $(|g\rangle+|e\rangle)/\sqrt{2} \leftrightarrow (|g\rangle-|e\rangle)/\sqrt{2}$. Damping results in a larger $|\mathrm{Tr}(\rho|gg\rangle\langle gg|)|$ component in the density matrix, while phase decoherence yields smaller off-diagonal terms in $\rho$. From the data at delay $t_{\rm d} = 400~\rm ns$, we estimate that $\sim 94\%$ percent of the infidelity is due to damping errors ($\sim 82\%$ from cable loss and $\sim 12\%$ from qubit decay) and $\sim 6\%$ percent is due to qubit decoherence.

\begin{figure*}
    \centering
    \includegraphics[width=12.9cm]{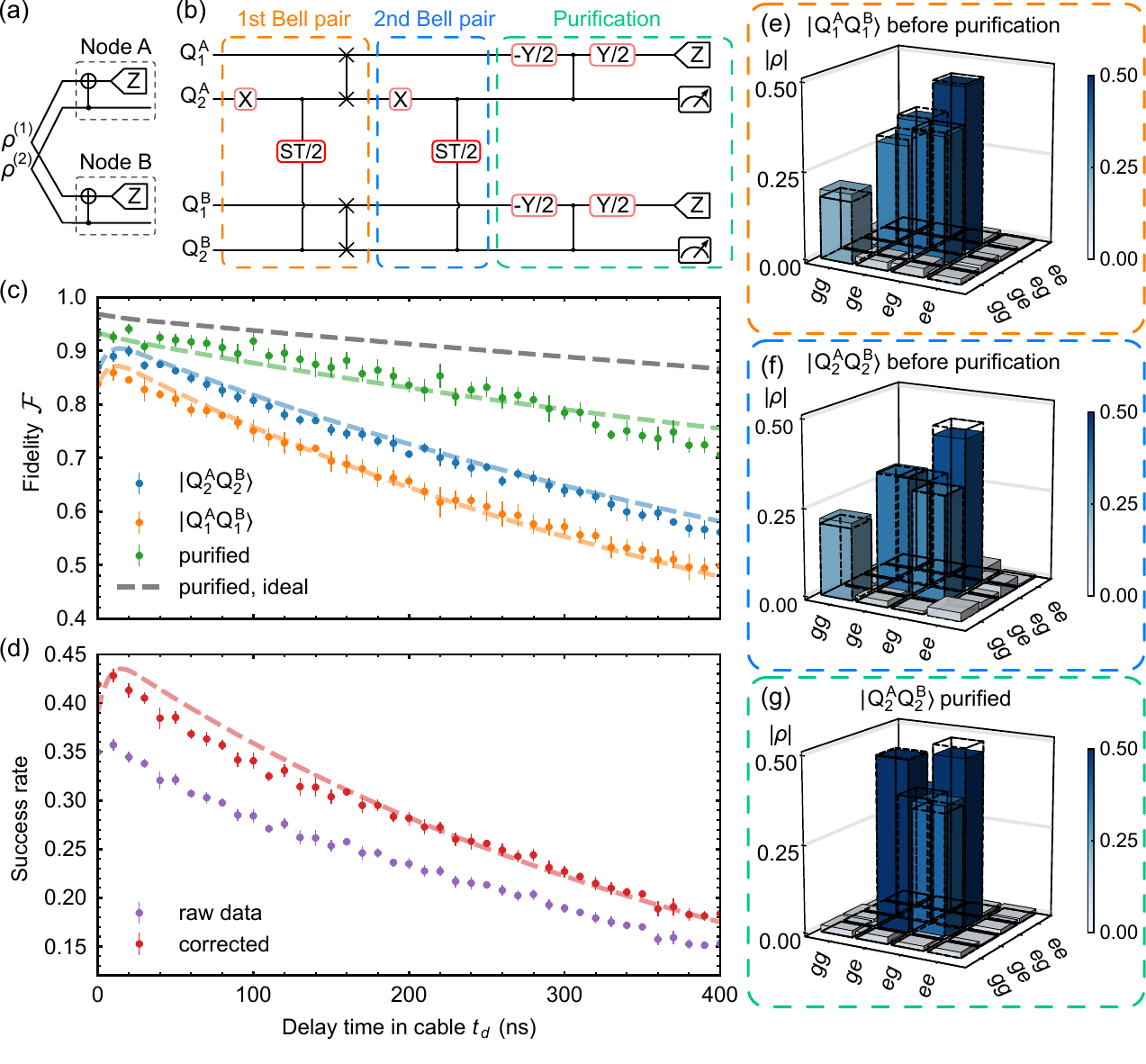}
    \caption{\label{fig:3} Entanglement purification. (a) Circuit schematic. (b) Experimental realization of the purification circuit in (a). The $\rm ST/2$ process is a ``half-photon'' transfer process as in Fig.~\ref{fig:2}(a). We prepare the first Bell pair $|Q_2^A Q_2^B\rangle$ followed by swaps into $|Q_1^A Q_1^B\rangle$. We then generate the second Bell pair in $|Q_2^A Q_2^B\rangle$, and purify using these two pairs. (c) Bell state fidelity $\mathcal{F}$ before purification for $|Q_2^A Q_2^B\rangle$ (blue) and $|Q_1^A Q_1^B\rangle$ (orange), and after purification for $|Q_2^A Q_2^B\rangle$ (green), each measured as a function of delay time $t_{\rm d}$. Grey dashed line is for error-free purification between two identical impure Bell states. (d) Success rate for purification, which is the probability of measuring $|Q_1^A Q_1^B\rangle\rangle$ in $|ee\rangle$ with (red) and without (purple) readout measurement correction \cite{Bialczak2010}. (e) State tomography with $t_{\rm{d}}=150~\rm ns$ for the pre-purification states $|Q_1^A Q_1^B\rangle$, representing $\rho^{(1)}$, with state fidelity $69.4\pm2.0\%$, and for (f) $|Q_2^A Q_2^B\rangle$, representing $\rho^{(2)}$, with state fidelity $75.3\pm1.0\%$. (g) Tomography for the post-purified $|Q_2^A Q_2^A\rangle$, representing $\rho_{\mathrm{f}}$, with state fidelity $86.9\pm 1.8\%$. Dashed lines are simulation results.}
\end{figure*}

We use a purification process including measurement and post-selection to improve the final Bell state fidelities, as shown in Fig.~\ref{fig:3}. The purification circuit is shown in Fig.~\ref{fig:3}(a), where two impure Bell pairs $\rho^{(1)}$ and $\rho^{(2)}$ are created between the two nodes, which serve as the source ($\rho^{(1)}$) and target pairs ($\rho^{(2)}$). Parallel CNOT gates are performed between the qubits in each node, followed by $Z$ measurements of the source pair $\rho^{(1)}$, using the measurement results to post-select from $\rho^{(2)}$. The result is a purified Bell pair $\rho_{\mathrm{f}}$ with a higher fidelity to the purification target state $|\psi^+\rangle = (|eg\rangle + |ge\rangle)/\sqrt{2}$ \cite{Bennett1996}.

Including only the errors due to the lossy channel and qubit dephasing, the two nominally-identical impure states can be written as
\begin{equation}
    \begin{split}
        \rho^{(1)} = \rho^{(2)} = (1-\epsilon_{p}) |\psi^-\rangle \langle\psi^-| + \epsilon_{p} |\psi^+\rangle \langle\psi^+| \\
        +\epsilon_{d}(|gg\rangle\langle gg|-|ge\rangle\langle ge|)
    \end{split}
\end{equation}
where $\epsilon_{d}$ accounts for any damping errors in the cable, of the form $|1\rangle \to |0\rangle$, and $\epsilon_{p}$ accounts for phase errors of the form $(|g\rangle+|e\rangle)/\sqrt{2} \leftrightarrow (|g\rangle-|e\rangle)/\sqrt{2}$, which take $|\psi^-\rangle$ to $|\psi^+\rangle = (|eg\rangle + |ge\rangle)/\sqrt{2}$.

When the two $Z$ measurements of $\rho^{(1)}$ are not consistent, the state $\rho_{\mathrm{f}}$ will be a mixed state, indicating a failed purification. Assuming perfect local operations and measurements, when the two measurement results are consistent, the result is a purified state closer to the ideal $|\psi^+\rangle$ pair. When the measurement result of $\rho^{(1)}$ is $|gg\rangle$, damping errors are partially corrected, with a non-zero ground state $|gg\rangle\langle gg|$ population. For more details see \cite{SupplementaryMaterial}.

When instead the measurement result is $|ee\rangle$, the final state will be
\begin{equation}
    \rho_{\mathrm{f}} = (1-\epsilon_{p_f})|\psi^+\rangle \langle\psi^+| + \epsilon_{p_f} |\psi^-\rangle \langle\psi^-|,
\end{equation}
with a phase error $\epsilon_{p_f} = (2\epsilon_{p}^2 - 2 \epsilon_{p} + 2 \epsilon_{d})/(1 - 2 \epsilon_{d})$ \cite{SupplementaryMaterial}. The damping error is fully corrected, with a purification success rate $0.5 - \epsilon_{d}$. Here we focus on the measurement result $|ee\rangle$, where the damping error is fully corrected and purification yields higher Bell state fidelities than the $|gg\rangle$ measurement result.

We implement the purification process as shown in Fig.~\ref{fig:3}(b). After generating the first Bell pair, shared between the two nodes in $|Q_2^A Q_2^B\rangle$, we apply two parallel $\mathrm{iSWAP}$ gates that transfer the state to $|Q_1^A Q_1^B\rangle$, with an efficiency over $99\%$. We then generate the second Bell pair in $|Q_2^A Q_2^B\rangle$ using the same sequence as for the the first Bell pair. We indirectly vary the cable loss by changing the delay time $t_{\mathrm{d}}$ the half-photon resides in the cable.

Pre-purification measurements of the two Bell states in $|Q_1^A Q_1^B\rangle$ and $|Q_2^A Q_2^B\rangle$, representing $\rho^{(1)}$ and $\rho^{(2)}$ respectively, are shown in Fig.~\ref{fig:3}(c), (e) and (f). These indicate the fidelity of the second Bell pair in $|Q_2^A Q_2^B\rangle$ is a few percent lower than the Bell pair in Fig.~\ref{fig:2}(b), due to imperfections in the $\mathrm{iSWAP}$ gates and possible interference with the first Bell pair during the second Bell pair generation. The first Bell pair's fidelity also falls due to qubit decay during the second Bell pair generation. Better qubit lifetimes \cite{Reagor2016,Place2021}, or the use of parallel communication channels, could reduce these infidelities. The purifying CNOT gate, with $|Q_2^A Q_2^B\rangle$ as the control, is implemented using a CZ gate combined with two single-qubit $Y/2$ gates applied to $|Q_1^A Q_1^B\rangle$. The CZ gate is realized using the qubits' second excited state $|f\rangle$, bringing the two-qubit states $|ee\rangle$ and $|gf\rangle$ into energy resonance so that the $|ee\rangle$ state acquires an extra phase compared to the other computational basis states \cite{Yamamoto2010}. We typically achieve CZ gate process fidelities of over $95\%$ \cite{SupplementaryMaterial}. Following the CNOT gates, we perform $Z$ measurements of $Q_1^{A,B}$ and tomography measurements of $Q_2^{A,B}$. We post-select as purified states those with $|Q_1^A Q_1^B\rangle = |ee\rangle$; this purification process targets the ideal Bell state $|\psi^+\rangle$.

\begin{figure}
    \centering
    \includegraphics[width=8.6cm]{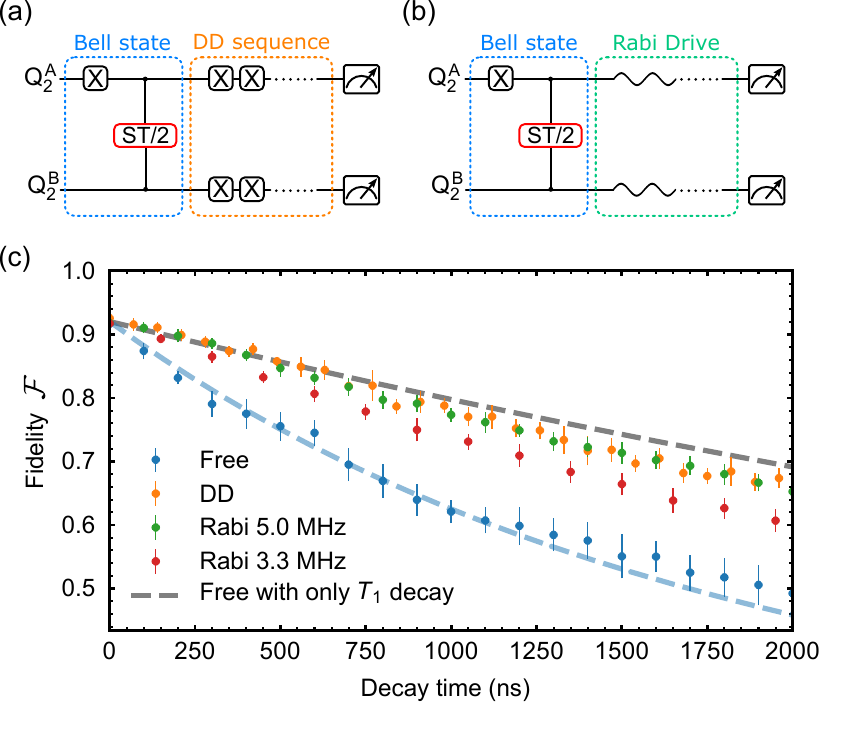}
    \caption{\label{fig:4}Entanglement protection using either dynamical decoupling (DD) or Rabi driving. (a) Pulse sequence for DD and (b) Rabi drive. The $\rm ST/2$ gate corresponds to the half-photon transfer process as shown in Fig.~\ref{fig:2}(b), with cable delay $t_{\rm d}=10~\rm ns$. (c) Bell state fidelity as a function of time for free evolution (blue), DD (orange), and for Rabi drive strengths $\Omega/2\pi=5~\rm MHz$ (green) and $\Omega/2\pi=3.3~\rm MHz$ (red). Numerical simulations are for free evolution including amplitude and phase decay (blue dashed line) and for free evolution with only $T_1$ decay (grey dashed line).}
\end{figure}

The fidelity of the purified state, representing $\rho_{\mathrm{f}}$, is shown in Fig.~\ref{fig:3}(c) as a function of delay $t_{\rm d}$. Larger $t_{\rm d}$ shows larger purification improvement, as there is more photon loss during cable transmission; the best fidelity of $94.09\%\pm 0.98\%$ is for the shortest delay $t_{\rm{d}}=20~\rm ns$. The largest fractional improvement in fidelity, defined as the change in fidelity divided by the initial (pre-purification) fidelity, is 25\%, achieved for the longest delay $t_{\rm d} = 400$~ns. The success rate, given by the probability of measuring $|Q_1^A Q_1^B\rangle$ in $|ee\rangle$, is shown in Fig.~\ref{fig:3}(d), which falls for longer delay times, as expected: The main limitation is due to storage decay of the first Bell pair, whose resultingly lower fidelity limits both the success rate and the purified fidelity. The gray dashed line shows the expected purified Bell state fidelity for two identical Bell pairs matching $|Q_2^AQ_2^B\rangle$. State tomography of the purified state for $t_{\rm d}=150~\rm ns$ is shown in Fig. \ref{fig:4}(g), with a state fidelity $86.9\pm 1.8\%$. The purified state has more than $10\%$ fidelity improvement and damping errors are mostly corrected.

The purification protocol is mostly limited by decoherence in the qubits. The dephasing time $T_2 \sim 3~\rm\mu s$ of our qubits is significantly shorter than the energy relaxation time $T_1 \sim 10~\mu s$, indicative of extra dephasing channels, possibly due to magnetic flux noise, to which frequency-tunable xmons are particularly susceptible \cite{Koch2007}. Using either dynamical decoupling (DD), or a simpler Rabi drive (RD), we can protect the Bell pairs from the local noise that generates some of this decoherence. DD is a technique commonly used in spin systems \cite{Carr1954, Meiboom1958}, where periodic pulse sequences average the effective environmental noise to near zero, yielding significantly extended qubit coherence times \cite{Bylander2011,Qiu2021,Pokharel2018}, as well as suppression of two-qubit correlated noise \cite{Chen2021}. The quantum circuit for DD is shown in Fig.~\ref{fig:4}(a), where we apply a sequence of $X$ gates to both qubits after generating a Bell state; the simpler RD is shown in Fig.~\ref{fig:4}(b). The DD $X$ gate we use is a $\pi$-pulse with an additional DRAG correction \cite{Motzoi2009}. The gate fidelity, as determined by randomized benchmarking \cite{Knill2008}, is $99.7\%$, with a gate duration of $30~\rm ns$ \cite{SupplementaryMaterial}. Following each $X$ gate, we insert $5~\rm ns$ of buffer time, so that each DD cycle, comprising two $X$ gates, takes $70~\rm ns$. To evaluate the performance of the DD sequence, we perform state tomography after a varying number of DD cycles, with the results shown in Fig.~\ref{fig:4}(d). We see that DD significantly improves the Bell state fidelity, approaching the fidelity associated with pure $T_1$ dephasing (gray dashed line). For a $1.4~\rm \mu s$ evolution time, the state fidelity improves from $57.6\pm3.0\%$ to $71.7\pm2.3\%$.

The simpler Rabi drive scheme works nearly as well as DD. The fidelity of the Bell pair with Rabi drive is shown in Fig.~\ref{fig:4}(d), showing similar performance to the DD sequence for $\Omega/2\pi = 5~\rm MHz$. For $1.4~\rm\mu s$ evolution, the RD fidelity improves from $57.6\pm3.0\%$ to $72.3\pm1.5\%$. For both DD and RD, the Bell state fidelity decay corresponds well to simulations using a qubit $T_2 \sim 12~\rm \mu s$, showing excellent protection of entanglement from local noise. We also tried to combine RD/DD with the purification protocol to further improve the performance. However this was not successful, we believe because of interference with the frequency-bias pulses applied to $|Q_1^AQ_1^B\rangle$ \cite{SupplementaryMaterial}.

In conclusion, we have demonstrated a two-node superconducting quantum network that supports the high-fidelity generation of Bell pairs across the network, with excellent state fidelity of $92.89\pm0.85\%$. Purification protocols successfully correct amplitude damping errors caused by the lossy communication channel, improving the state fidelity to $94.09\pm 0.98\%$. Furthermore, local phase decoherence can be minimized using either dynamical decoupling or the simpler Rabi drive. These results point to the powerful potential for distributed quantum computing in superconducting networks, relying on high-fidelity entanglement over meter-scale networks.

\section{Acknowledgements}
\begin{acknowledgments}
We thank Liang Jiang, David Schuster and Peter Duda for helpful discussions, and W. D. Oliver and G. Calusine at MIT Lincoln Lab for providing the traveling-wave parametric amplifier (TWPA) used in this work. Devices and experiments were supported by the Air Force Office of Scientific Research and the Army Research Laboratory. \'E.D. was supported by LDRD funds from Argonne National Laboratory; A.N.C. was supported in part by the DOE, Office of Basic Energy Sciences. This work was partially supported by UChicago's MRSEC (NSF award DMR-2011854) and by the NSF QLCI for HQAN (NSF Award 2016136). We made use of the Pritzker Nanofabrication Facility, which receives support from SHyNE, a node of the National Science Foundation's National Nanotechnology Coordinated Infrastructure (NSF Grant No. NNCI ECCS-2025633). The authors declare no competing financial interests. Correspondence and requests for materials should be addressed to A. N. Cleland (anc@uchicago.edu).
\end{acknowledgments}

\end{document}


\begin{center}
\textbf{\large Supplementary Material for ``Entanglement purification and protection in a superconducting quantum network''}
\end{center}
\setcounter{equation}{0}
\setcounter{figure}{0}
\setcounter{table}{0}
\setcounter{page}{1}
\makeatletter
\renewcommand{\theequation}{S\arabic{equation}}
\renewcommand{\thefigure}{S\arabic{figure}}
\renewcommand{\thetable}{S\arabic{table}}
\renewcommand{\bibnumfmt}[1]{[S#1]}
\renewcommand{\citenumfont}[1]{S#1}
\setcounter{figure}{0}

\section{Device and Experimental Setup}
\begin{figure}[ht]
    \centering
    \includegraphics[width=0.8\textwidth]{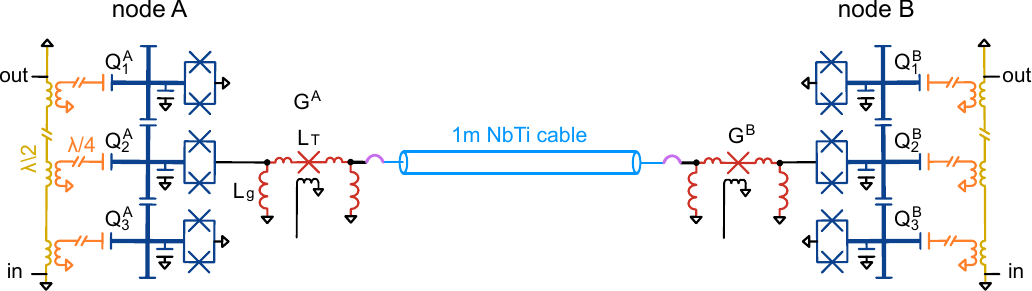}
    \caption{Circuit diagram for two-node network. More details can be found in Ref. \cite{Zhong2021}.}\label{fig:S1}
\end{figure}

The device and setup used in this experiment are identical to those used in Ref. \cite{Zhong2021} except that the qubit operating points are different in this experiment. The simplified circuit diagram of the device is shown in Fig.~\ref{fig:S1}. The center qubit $Q_2^{A,B}$ in each node is directly wire-bonded to the 1-m long superconducting NbTi coaxial cable with a tunable inductive coupler $G^{A,B}$ \cite{Chen2014}. The coupling strength $g^{A,B}$ is tuned with a dedicated flux line. More details on the fabrication and measurement system can be found in Ref. \cite{Zhong2021}. The parameters of the qubits are listed in Section \ref{qubitchar}.

\section{Qubit and Cable Characterization}\label{qubitchar}
The basic qubit parameters are shown in Table \ref{tab:S1}. The inductance of the coupler $G^k$ contributes to the total inductance of the qubit $Q_2^k$, which decreases $Q_2^k$'s anharmonicity. $T_\phi$ is determined by a Ramsey measurement. As qubits $Q_2^k$ operate near their maximum frequency points, these qubits have longer dephasing time than the other qubits $Q_{1,3}^k$.  To achieve fast readout of qubits, we include a Purcell filter \cite{Jeffrey2014} between the readout resonators and the readout line in each node. The Purcell filter has a resonant frequency of around $6.5~\rm GHz$, protecting the qubit from decay through the readout resonator. In node $A$, we use a traveling-wave parametric amplifier (TWPA) \cite{Macklin2015} as the first-stage amplifier for the readout signal. The resulting average readout visibilities of the qubit $|g\rangle$ and $|e\rangle$ states are around $95\%$, limited by the qubit lifetime and residual excited-state population.

\begin{table}
    \begin{tabular}{|c|c|c|c|c|c|c|c|c|c|}
    \hline
            & $f_{eg}^{\mathrm{max}}\mathrm{(GHz)}$ & $f_{eg}\mathrm{(GHz)}$ & $\eta\mathrm{(GHz)}$ & $T_1\mathrm{(\mu s)}$ & $T_\phi\mathrm{(\mu s)}$ & $f_{rr}\mathrm{(GHz)}$ & $\tau_{rr}\mathrm{(ns)}$ & $F_g$ & $F_e$ \\ \hline
    $Q_1^A$ & 6.03                                  & 5.2556                                 & -0.23                & 8.9                   & 1.8                      & 6.6050                 & 350                      & 0.968 & 0.940 \\
    $Q_2^A$ & 6.14                                  & 5.8695                                 & -0.15                & 5.7                   & 3.1                      & 6.5497                 & 450                      & 0.974 & 0.927 \\
    $Q_3^A$ & 6.04                                  & 5.5055                                 & -0.23                & 6.3                   & 2.5                      & 6.5032                 & 300                      & 0.962 & 0.926 \\
    $Q_1^B$ & 6.08                                  & 5.3021                                 & -0.23                & 22.1                  & 2.2                      & 6.5056                 & 400                      & 0.988 & 0.936 \\
    $Q_2^B$ & 6.25                                  & 5.8901                                 & -0.15                & 9.2                   & 3.0                      & 6.5560                 & 600                      & 0.965 & 0.939 \\
    $Q_3^B$ & 6.16                                  & 5.3218                                 & -0.23                & 21.1                  & 1.8                      & 6.6080                 & 350                      & 0.983 & 0.939 \\ \hline
    \end{tabular}\caption{\label{tab:S1}Qubit parameters. The maximum qubit frequency is $f_{eg}^{\mathrm{max}}$ and $f_{eg}$ is the qubit operating frequency used here; $\eta$ is the qubit anharmonicity; $T_1$ and $T_\phi$ are the qubit lifetime and pure dephasing time at the operating point; $f_{rr}$ is the readout resonator frequency; $\tau_{rr}$ is the readout pulse duration; and $F_g$ and $F_e$ are the readout visibilities of $|g\rangle$ and $|e\rangle$ states, respectively.}
\end{table}

We use randomized benchmarking (RB) to characterize the single qubit gate performance \cite{Knill2008}. Our single-qubit $\pi$ gate duration is $30~\rm ns$ and $\pi/2$ gate duration is $20~\rm ns$, for all rotation axes, with gate durations optimized to balance qubit lifetime with state leakage. To minimize the effect of the second excited state $|f\rangle$, we use a DRAG correction for all our single qubit gates \cite{Motzoi2009}. We get an average single qubit RB gate fidelity of $99.8\%$.

We tune the frequency of our qubits to implement $\rm iSWAP$ and $\rm CZ$ gates. To swap an excitation between $Q_1^k$ and $Q_2^k$ in the same node, we bring $Q_1^k$ and $Q_2^k$ on resonance. At $\tau_{\rm iSWAP}=\pi/2g_{12}^k\approx 15~\rm ns$ ($k=A,B$), we complete the $|eg\rangle\to -i|ge\rangle$ process, which forms a $\rm iSWAP$ gate. To measure the efficiency of the $\rm iSWAP$ gate, we first rotate $Q_2^k$ to $|e\rangle$ and measure the excited state population $P_e$. Next, with $Q_2^k$ in $|g\rangle$, we prepare $Q_1^k$ in $|e\rangle$, then swap the excitation to $Q_2^k$, followed by a measurement of $P_e$ for $Q_2^k$. Comparing the resulting $P_e$ in the two experiments, we find a swap efficiency of $\sim 99\%$.

\begin{figure}
    \centering
    \includegraphics[width=0.8\textwidth]{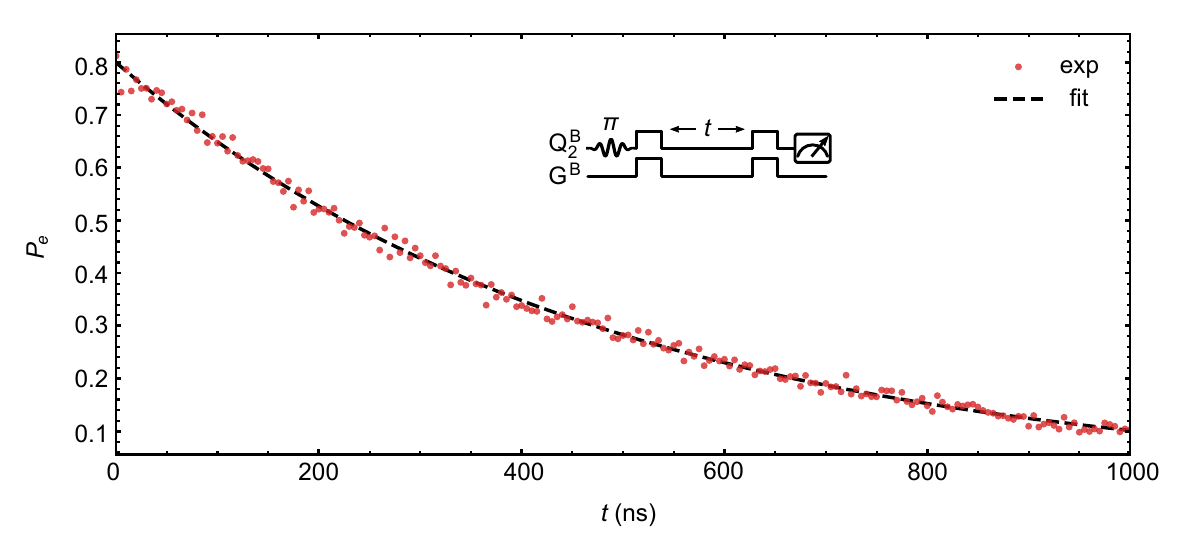}
    \caption{\label{fig:S2}$T_{1r}$ measurement of the $5.806~\rm GHz$ cable mode. We prepare $Q_2^B$ in $|e\rangle$ and swap the excitation to the cable mode, wait for a time delay $t$, then swap the excitation back to the qubit and measure the excited state population $P_e$. The swap duration is fixed to $30~\rm ns$. The measured $T_{1r}$ is $477.3\pm 8.1~\rm ns$.}
\end{figure}

\begin{figure}
    \includegraphics[width=0.8\textwidth]{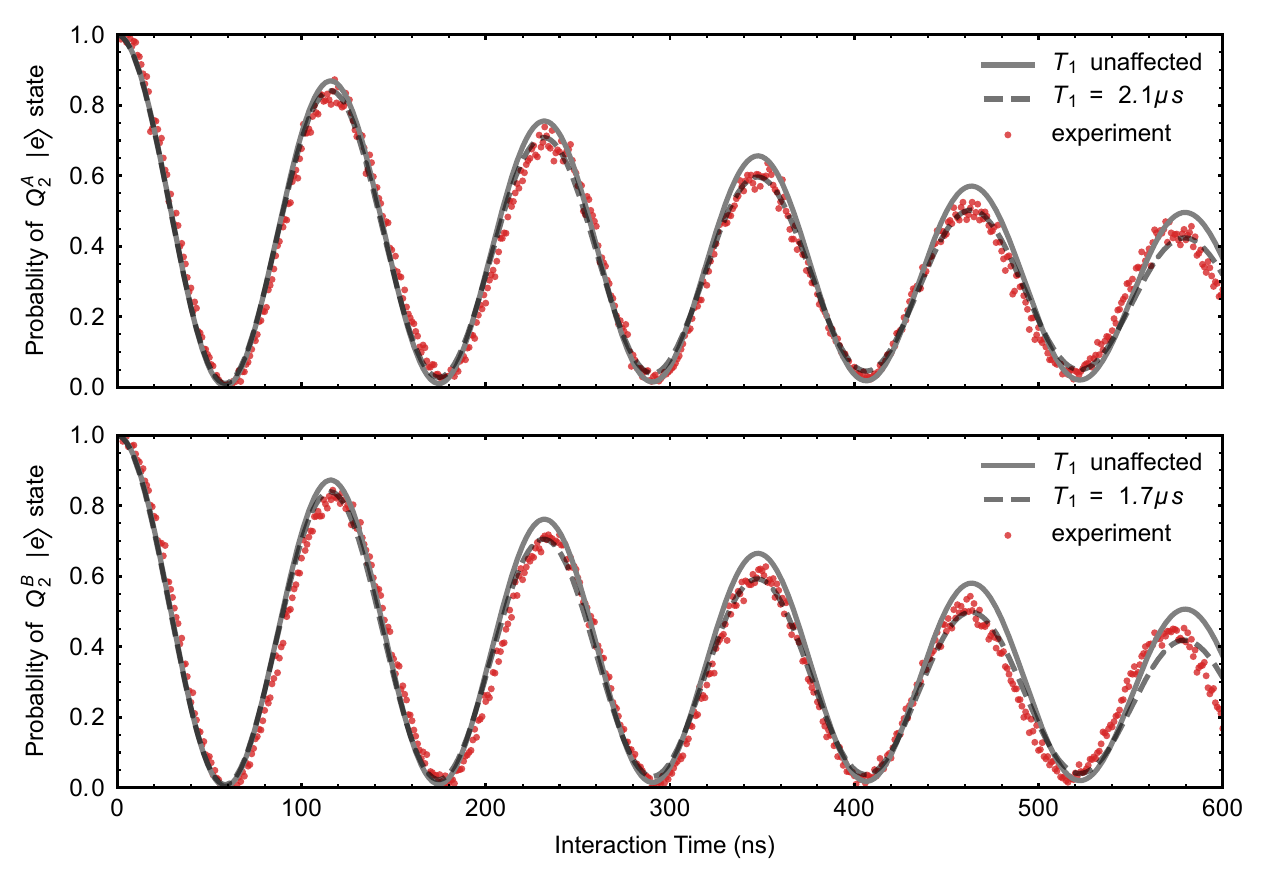}
    \caption{Vacuum Rabi oscillation between $5.806~\rm GHz$ mode and $Q_2^A$ (top panel) and $Q_2^B$ (bottom panel). Red circles are experimental data, gray lines are numerical simulations with $T_1$ as shown in Table \ref{tab:S1}. Grey dashed lines are numerical simulations with shortened $T_1$, which implies that $T_1$ is affected by the coupling strength between the qubit and cable \cite{Zhong2021}.}\label{fig:S3}
\end{figure}

To implement a $\rm CZ$ gate between $Q_1^k$ and $Q_2^k$, we bring $|ee\rangle$ and $|gf\rangle$ on resonance. When $|ee\rangle$ is on resonance with $|gf\rangle$, the two-qubit state will evolve as $|ee\rangle \to -i|gf\rangle \to -|ee\rangle$, yielding a CZ gate in a time $\tau_{\rm CZ} \approx 21\rm ns$. To characterize the gate performance, we perform process tomography \cite{Yamamoto2010} to obtain the process $\chi$ matrix of the $\rm CZ$ gate. To overcome the issue of dynamical phase accumulated during the $\rm CZ$ gate, here we directly tune the phase of the tomography pulses to get the right $\rm CZ$ gate process $\chi$ matrix. We find process fidelities $\mathcal{F}_p = \mathrm{Tr}(\chi_{\rm ideal}\cdot\chi_{\rm exp})$ for the $\rm CZ$ gate between $Q_1^A-Q_2^A$ and $Q_1^B-Q_2^B$ of $95.93\%\pm0.64\%$ and $95.05\%\pm0.68\%$, respectively. With two $\pi/2$ single qubit gates and a $\rm CZ$ gate, we can implement a $\rm CNOT$ gate. To overcome the dynamical phase problem in experiment, we fix one $\pi/2$ gate phase and tune the phase of the other $\pi/2$ gate. The $\rm CNOT$ gate in the experiment will introduce additional phase differences between the two qubits, but this is unimportant for the purification experiment, where we only correct damping errors.

We use the center qubits $Q_2^k$ to characterize the communication mode in the cable. Each qubit is coupled to the cable with a tunable inductive coupler $G^k$. The effective coupling strength $g_n$ between the qubit and the $n$th Fabry-P\'erot mode in the cable is given by \cite{Zhong2019}
\begin{equation}
    g_n = \frac{\sqrt{\omega_q\omega_n}}{2}\frac{M}{\sqrt{(L_{J}+L_g)(L_{n}+L_g)}},
\end{equation}
where $\omega_q$ is the qubit frequency, $\omega_n$ is the $n$th mode frequency, $L_J$ is the qubit inductance, $L_n$ is inductance of the lumped element of the $n$th mode ($L_n \sim 121~\rm nH$ for the $5.806~\rm GHz$ mode, with $n \approx 55$), and $L_g = 0.2~\rm nH$ is the inductance to ground (see Fig.~\ref{fig:S1}). $M$ is the effective mutual inductance between the qubit and the communication mode,
\begin{equation}
        M = \frac{L_g^2}{2L_g + L_w + L_{T}/\cos\delta},
\end{equation}
where $\delta$ is the flux-tunable phase across the coupler Josephson junction, $L_T\approx 0.62~\rm nH$ is the coupler inductance at $\delta=0$, and $L_w\approx 0.1~\rm nH$ is the stray wiring inductance is series with $L_T$. The maximum coupling strength achieved by tuning $\delta$ is $g_{\rm max}/2\pi \approx 28~\rm MHz$. The qubit frequency is slightly shifted when the coupling is turned on, which is compensated using a qubit $Z$ pulse. Here we focus on the $5.806~\rm GHz$ communication mode and only turn on the coupler during the Bell state generation. To characterize this mode, we first prepare $Q_2^B$ in $|e\rangle$ and swap the excitation to the cable, leave the excitation in the cable mode for a time $t$ and then swapping it back to $Q_2^B$ to measure the remnant excitation, shown in Fig.~\ref{fig:S2}. The swap duration used for this measurement is $30~\rm ns$. The measured $T_{1r}$ is $477.3\pm 8.1~\rm ns$, corresponding to a quality factor $Q = 1.7\times 10^4$. To characterize the performance when the coupler is turned on, we measure the vacuum Rabi oscillations between the cable mode and qubit, shown in Fig.~ \ref{fig:S3}. Numerical simulations show that the qubit $T_1$ and $T_\phi$ are shortened to around $2~\rm \mu s$ and $1~\rm \mu s$ respectively, likely due to losses associated with the wire-bond connection \cite{Zhong2021}.

\section{State Tomography and Readout Correction}
The density matrices are characterized using quantum state tomography \cite{Steffen2006}, where we apply gates from the set $\{I, X/2, Y/2\}$ to each qubit after state generation, and perform simultaneous readout of all qubits. Density matrices are reconstructed from the nine measured probabilities and are constrained to be Hermitian and unitary.

To perform the process tomography \cite{Yamamoto2010}, we first prepare different initial states by applying gates from the set $\{I, X, X/2, Y/2\}$ to each qubit, perform the target process (i.e. CZ gate, state transfer) and do state tomography to determine the process $\chi$ matrix.

The single-shot measurements are repeated $8\times 10^3$ times to extract probabilities. The measured probabilities $\mathbf{P^M}=(P_g^M,P_e^M)^T$ are corrected for readout errors using the pre-measured readout visibility data ($F_g$, $F_e$) \cite{Bialczak2010}. The state probabilities $\mathbf{P}=(P_g,P_e)^T$ are reconstructed by $\mathbf{P}=\mathbf{F}^{-1}\mathbf{P^M}$, where
\begin{equation}
  \mathbf{F}=\begin{bmatrix}
    F_g&1-F_e\\1-F_g&F_e
  \end{bmatrix}.
\end{equation}
Each measurement is repeated independently $20$ times to estimate the overall uncertainties, defined as the standard deviation of these $20$ measurements.

\section{Numerical Simulations}
The full Hamiltonian of our system in the rotating frame can be written using the following multi-qubit, multi-mode communication channel model:
\begin{equation}
    \begin{split}
            H/\hbar=&\sum_{i=1,2,3}^{k=A,B}\Delta\omega_i^k\sigma^{k\dagger}_i\sigma^{k}_i+\sum_{m=1}^M\left(m-\frac{M+1}{2}\right)\omega_{\mathrm{FSR}}b^\dagger_mb_m\\
            &+\sum_{j=1,3}^{k=A,B}g_{j,2}^k(\sigma_2^k\sigma_j^{k\dagger}+\sigma_{2}^{k\dagger}\sigma_j^k)\\
            &+\sum_{m=1}^{M}g^A_m(\sigma_2^Ab_m^\dagger+\sigma_2^{A\dagger}b_m)+\sum_{m=1}^M(-1)^mg^B_m(\sigma_2^Bb_m^\dagger+\sigma_2^{B\dagger}b_m),
    \end{split}
\end{equation}
where $\sigma_i^k$ and $b_m$ are the annihilation operators for qubit $Q_i^k$ and the $m$th cable mode respectively, $\Delta\omega_i^k$ is the qubit frequency detuning with respect to the rotating frame frequency, $M$ is the number of standing modes considered in the simulation, $\omega_{\mathrm{FSR}}$ is free spectral range of the cable modes, $g_{j,2}^k$ is the coupling strength between $Q_j^{k}$ and $Q_2^k$, and $g_m^{k}$ is the coupling strength between $Q_2^k$ and the $m$th cable mode. The rotating frame frequency is set to the frequency of the communication mode. The sign of $g^B_m$ alternates with the mode number $m$ due to the parity dependence of the standing wave modes.

In Fig. 1(b), only $Q_2^A$ interacts with the communication mode, and is in resonance with that mode. The Hamiltonian can then be simplified to
\begin{equation}
    H/\hbar=\sum_{m=1}^M\left(m-\frac{M+1}{2}\right)\omega_{\mathrm{FSR}}b^\dagger_mb_m+\sum_{m=1}^{M}g^A_m(\sigma_2^Ab_m^\dagger+\sigma_2^{A\dagger}b_m)
\end{equation}
where we set $M=3$ and constrain to a single excitation in the system. The master equation simulation is then compared to the experimental data, where the cable mode lifetime $T_{1r}$ is as characterized in Fig. \ref{fig:S2}, but where we find that the qubit lifetime $T_1$ needs to be changed to $2.1~\mathrm{\mu s}$ to better fit the experimental data (Fig. \ref{fig:S3}). We use the pure dephasing time $T_{\phi}$ shown in Table \ref{tab:S1} in the simulations.

In Fig. 2(a-d), $Q_1^k$ and $Q_3^k$ are detuned away from $Q_{2}^k$. The time-dependent Hamiltonian can be simplified to
\begin{equation}\label{equ:S6}
    \begin{split}
            H(t)/\hbar=&\sum_{k=A,B}\Delta\omega_2^k(t)\sigma^{k\dagger}_2\sigma^{k}_2+\sum_{m=1}^M\left(m-\frac{M+1}{2}\right)\omega_{\mathrm{FSR}}b^\dagger_mb_m\\
            &+\sum_{m=1}^{M}g^A_m(t)(\sigma_2^Ab_m^\dagger+\sigma_2^{A\dagger}b_m)+\sum_{m=1}^M(-1)^mg^B_m(t)(\sigma_2^Bb_m^\dagger+\sigma_2^{B\dagger}b_m)
    \end{split}
\end{equation}
where we set $M=3$ and constrain to a single excitation in the system. We assume perfect rectangular pulses for the frequency changes and coupler modulations. The qubit lifetime $T_1$ is taken from fits to the vacuum Rabi oscillations (Fig. \ref{fig:S3}) when the coupler is on, and from Table \ref{tab:S1} when the coupler is off; the pure dephasing time $T_\phi$ is from Table \ref{tab:S1}, and the cable mode lifetime is from Fig. \ref{fig:S2}.

In Fig. 3(c-g), the Bell states are simulated using Eq. \ref{equ:S6}, for which we assume perfect local operations. The first Bell pair decays during the second Bell pair generation, where we directly use $Q_1^k$'s parameters in Table \ref{tab:S1}. In Fig. 4(c), the initial Bell state is also simulated using Eq. \ref{equ:S6} above. We vary the qubit pure dephasing time $T_\phi$ and fix qubit $T_1$ to fit the simulations to the experimental data.

\section{Discussion of Entanglement Purification}
The four Bell states $|\psi^{\pm}\rangle$, $|\phi^{\pm}\rangle$ are
\begin{align}
  |\psi^{\pm}\rangle&=\frac{1}{\sqrt{2}}(|eg\rangle\pm|eg\rangle)\\
  |\phi^{\pm}\rangle&=\frac{1}{\sqrt{2}}(|gg\rangle\pm|ee\rangle)
\end{align}
The ideal Bell state via cable-mediated entanglement is $|\psi^-\rangle$,
\begin{equation}
    \rho_{\rm ideal} = |\psi^-\rangle\langle\psi^-| = \frac{1}{2} \begin{bmatrix} 0&0&0&0\\0&1&-1&0\\0&-1&1&0\\0&0&0&0 \end{bmatrix}.
\end{equation}
For a bit-flip error channel with error probability $p$, the process is represented by the operators
\begin{align}
    E_0 &= \sqrt{1-p} I, \\
    E_1 &= \sqrt{p} X.
\end{align}
If we assume only one qubit suffers from a bit-flip error, the final state will be
\begin{equation}
    \rho = \sum_i I\otimes E_i \cdot \rho_{\rm ideal} \cdot I \otimes E_i^\dagger = (1-\frac{p}{2}) |\psi^-\rangle \langle\psi^-| + \frac{p}{2} |\phi^- \rangle \langle\phi^-|,
\end{equation}
with a fidelity $\mathcal{F} = 1-p/2$. After applying the bit-purification circuit shown in Fig.~\ref{fig:S6}(a), the final state fidelity will be
\begin{equation}
  \mathcal{F}' = \frac{\mathcal{F}^2}{\mathcal{F}^2+(1-\mathcal{F})^2},
\end{equation}
with post-selection of $|gg\rangle$ or $|ee\rangle$.

For a phase-flip error channel, the process is represented by
\begin{align}
    E_0 &= \sqrt{1-p} I ,\\
    E_1 &= \sqrt{p} Z.
\end{align}
If we again assume only one qubit suffers from a phase-flip error, the final state will be
\begin{equation}
    \rho = \sum_i I \otimes E_i \cdot \rho_{\rm ideal} \cdot I \otimes E_i^\dagger = (1-\frac{p}{2}) |\psi^-\rangle \langle\psi^-| + \frac{p}{2} |\psi^+\rangle \langle\psi^+|,
\end{equation}
with a final state fidelity $\mathcal{F} = 1-p/2$, where $|\psi^+\rangle = (|eg\rangle + |ge\rangle)/\sqrt{2}$. The phase-purification circuit shown in in Fig.~\ref{fig:S6}\textbf{(b)} acts in similar fashion to the bit-flip purification circuit.

Considering an amplitude-damping error channel, the channel representation is
\begin{align}
    E_0 &= \begin{bmatrix} 1 & 0 \\ 0&\sqrt{1-p} \end{bmatrix}, \\
    E_1 &= \begin{bmatrix} 0 & \sqrt{p} \\ 0 & 0 \end{bmatrix}.
\end{align}
Assuming only one qubit suffers an amplitude-damping error, the final state will be
\begin{equation}
    \begin{split}
        \rho &= \sum_i I \otimes E_i \cdot \rho_{\rm ideal} \cdot I \otimes E_i^\dagger \\
        &= \frac{1+\sqrt{1-p}}{2} |\psi^-\rangle \langle\psi^-| + \frac{1-\sqrt{1-p}}{2} |\psi^+\rangle \langle\psi^+| + \frac{p}{2} (|gg\rangle \langle gg| - |ge\rangle \langle ge|) \\
        &= \frac{1}{2} \begin{bmatrix} p & 0 & 0 & 0 \\ 0 & 1 - p &- \sqrt{1-p} & 0 \\ 0 & -\sqrt{1-p} & 1 & 0 \\ 0 & 0 & 0 & 0 \end{bmatrix},
    \end{split}
\end{equation}
with fidelity
\begin{equation}
  \mathcal{F} = \frac{2-p+2\sqrt{1-p}}{4} \sim 1-\frac{p}{2}.
\end{equation}
Applying the bit-purification circuit with post-selection of the $|gg\rangle$ state, the final state will be
\begin{align}
  \rho' &= \frac{1}{2(1-p+p^2)}\begin{bmatrix} p^2 & 0 & 0 & 0 \\ 0 & (1-p)^2 &1-p & 0 \\ 0 & 1-p & 1 & 0 \\ 0 & 0 & 0 & 0 \end{bmatrix}, \\
  \mathcal{F}' &= \frac{(2-p)^2}{4(1-p+p^2)}\sim1-\frac{3p^2}{4},
\end{align}
With a success rate $(1-p+p^2)/2$ and only part of damping errors are corrected. If we do the post-selection with the $|ee\rangle$ state, the final state will be
\begin{align}
    \rho' &= |\psi^+\rangle \langle\psi^+|, \\
    \mathcal{F}' &= 1,
\end{align}
with a success rate of $(1-p)/2$ and correcting all the damping errors.

If we consider both phase and amplitude-damping errors, the state before purification can be written as
\begin{align}
    \rho &= (1-\epsilon_{p}) |\psi^-\rangle \langle\psi^-| + \epsilon_{p} |\psi^+\rangle \langle\psi^+| + \epsilon_{d} (|gg\rangle \langle gg| - |ge\rangle \langle ge|), \\
    \mathcal{F} &= 1 - \frac{\epsilon_{d}}{2} - \epsilon_{p}.
\end{align}
After applying the bit-purification circuit with post-selection of $|gg\rangle$, the final state will be
\begin{align}
  \rho'&=\frac{1}{2(1-2\epsilon_d+4\epsilon_d^2)}\begin{bmatrix} 4\epsilon_d^2 & 0 & 0 & 0 \\ 0 & (1-2\epsilon_d)^2 &(1-2\epsilon_p)^2 & 0 \\ 0 & (1-2\epsilon_p)^2 & 1 & 0 \\ 0 & 0 & 0 & 0 \end{bmatrix}, \\
  \mathcal{F}'&=\frac{1-\epsilon_d-2\epsilon_p+\epsilon_d^2+2\epsilon_p^2}{1-2\epsilon_d+4\epsilon_d^2}
\end{align}
with a success rate $0.5-\epsilon_d+2\epsilon_d^2$ and non-zero ground state population $2\epsilon_d^2/(1-2\epsilon_d+4\epsilon_d^2)$. If we do the post-selection of $|ee\rangle$, the final state will be
\begin{align}
  \rho' &= (1-\epsilon_{p}') |\psi^+\rangle \langle\psi^+| + \epsilon_{p}' |\psi^-\rangle \langle\psi^-|, \\
  \epsilon_{p}' &= \frac{2\epsilon_{p}-2\epsilon_{p}^2-\epsilon_{d}}{1-2\epsilon_{d}}, \\
  \mathcal{F}' &= 1 - \epsilon_{p}' \cong 1+\epsilon_{d}-2\epsilon_{p},
\end{align}
with a success rate $0.5-\epsilon_{d}$ and zero ground state population.

\begin{figure}
    \centering
    \includegraphics[width=0.8\textwidth]{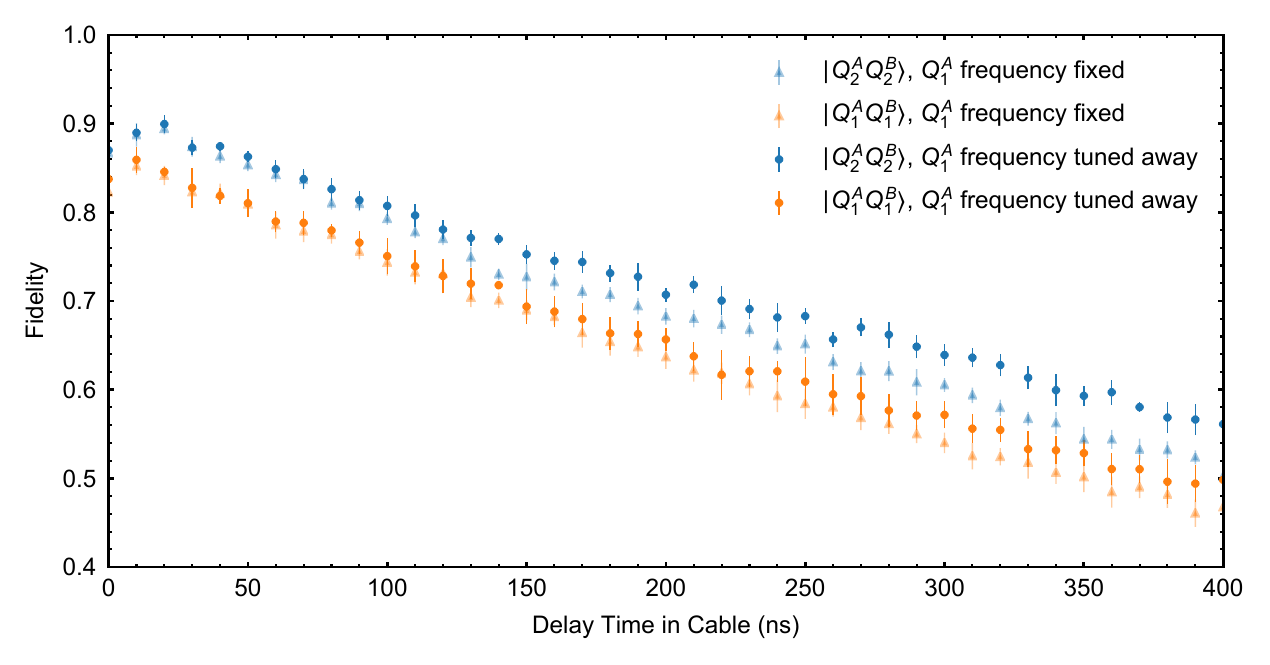}
    \caption{Bell state fidelity before purification with (circles) and without (triangles) tuning away $Q_1^A$'s operating frequency. The fidelity decreases if $Q_1^A$'s frequency is left fixed.}\label{fig:S4}
\end{figure}

\begin{figure}
    \centering
    \includegraphics[width=0.8\textwidth]{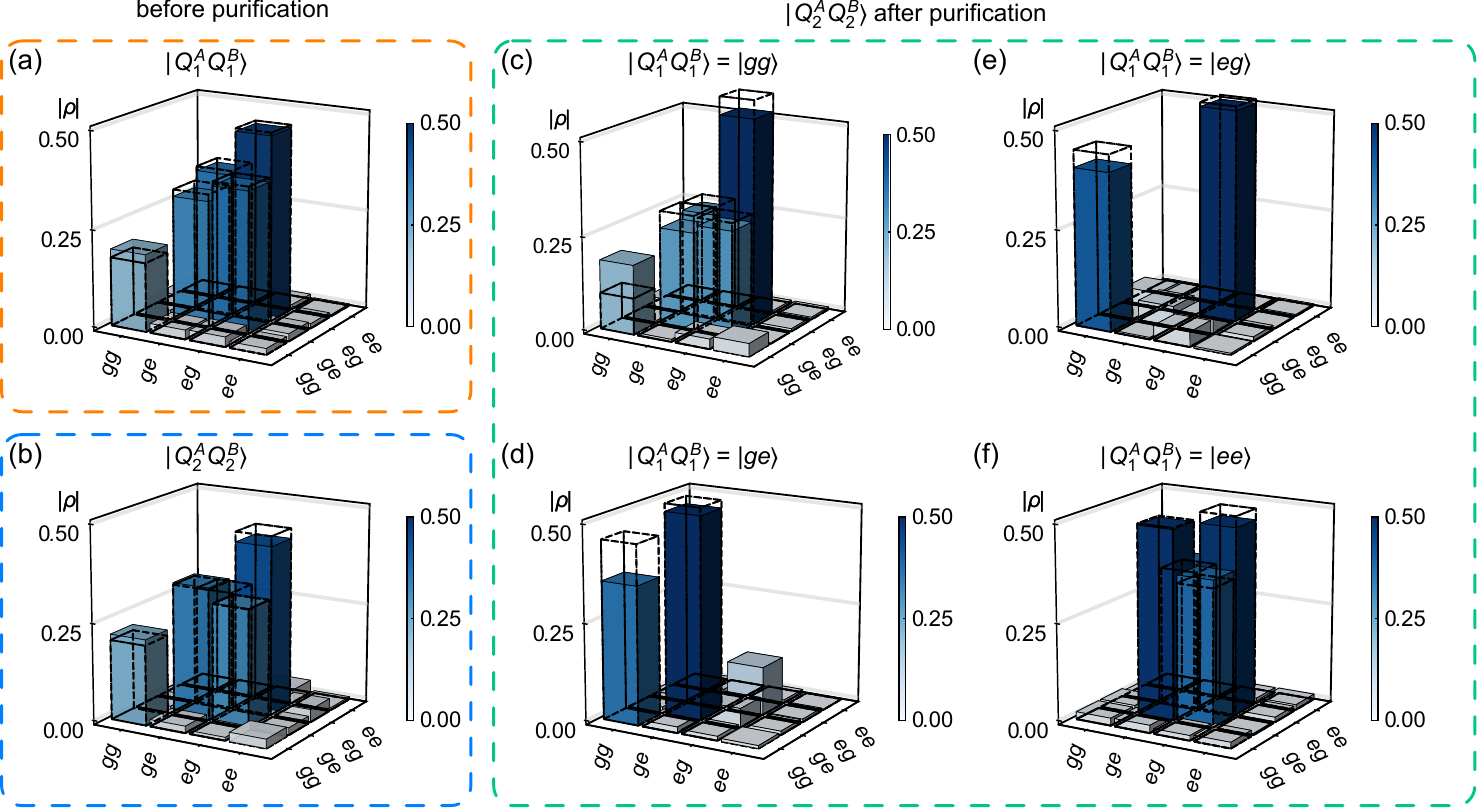}
    \caption{Full state tomography of entanglement purification results when $t_d = 150~\rm ns$. (a-b) State tomography of (a) $|Q_1^A Q_1^B \rangle$ and (b) $|Q_2^A Q_2^B \rangle$ before purification. (c-f) State tomography of $|Q_2^A Q_2^B \rangle$ for different $|Q_1^A Q_1^B \rangle$ measurement results. Dashed lines are numerical simulation results.}\label{fig:S5}
\end{figure}

To realize entanglement purification, we need to prepare at least two Bell pairs. Since there is only one channel in our system, this is done sequentially, first creating and then storing the first Bell pair in the storage qubits $Q_1^k$. In the experiment, we find that the Bell state fidelity suffers if we leave the qubit frequency of $Q_1^A$ fixed (see Fig.~\ref{fig:S4}). This could be due to two-photon processes between $Q_1^A$ and $Q_2^A$ during the excitation pulse used for generating the second Bell state, as when generating the second Bell pair, both $Q_1^A$ and $Q_2^A$ have a population in their $|e\rangle$ states. We find that when we tune $Q_1^A$'s $|g\rangle-|e\rangle$ frequency from $5.25~\rm GHz$ to $\sim 4.9~\rm GHz$, the fidelity of the second Bell state is closer to that of the initial state fidelity.

For entanglement purification, we perform full state tomography of all the final states (Fig.~\ref{fig:S5}). As is discussed in the main text, when the measurement results for $|Q_1^A Q_1^B\rangle$ are not consistent, the final state of $|Q_2^A Q_2^B \rangle$ will be a mixed state. When $|Q_1^A Q_1^B \rangle = |gg\rangle$, the final state of $|Q_2^A Q_2^B\rangle$ will have both damping and phase errors. Only when $|Q_1^A Q_1^B \rangle = |ee\rangle$, then the damping error is corrected, consistent with the experimental measurements.

\begin{figure}
    \centering
    \includegraphics[width=0.8\textwidth]{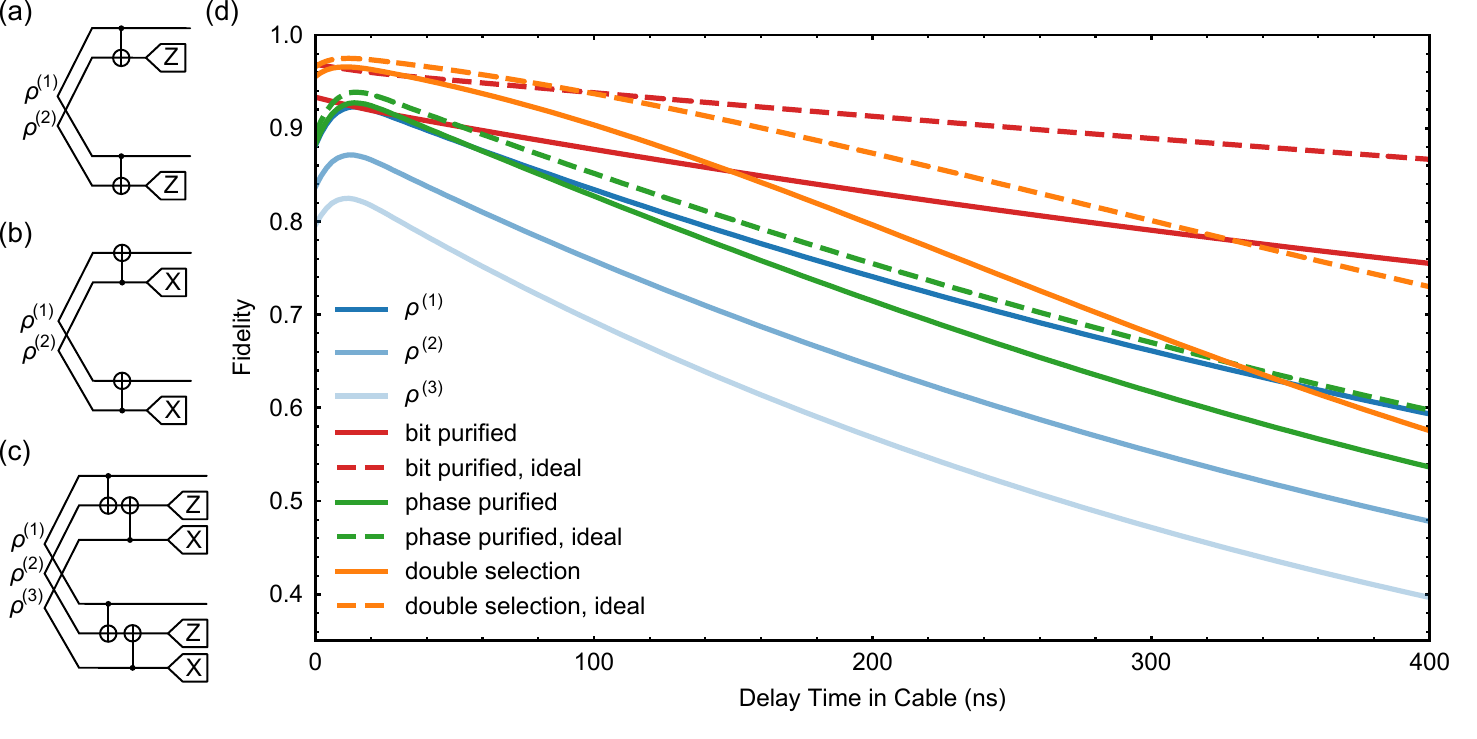}
    \caption{Results of numerical simulations to evaluate different entanglement purification protocols. (a-c) Different entanglement purification circuits used in numerical simulations: (a) bit-error purification circuit, (b) phase-error purification circuit, (c) double-selection purification circuit \cite{Fujii2009}. (d) Simulated purification results for the protocols in (a-c). Blue lines are states before purification including (lightest) and not including (darkest) state decay during state storage. Simulated results are for the bit-purification (red), phase-purification (green) and purification with double selection (orange). Dashed lines are purification results with ideal Bell pairs $\rho^{(0)}$, while solid lines include state decay during storage.}\label{fig:S6}
\end{figure}

There are a number of other entanglement purification protocols \cite{Bennett1996,Jiang2007,Fujii2009} which can be implemented using this system. We performed numerical simulations to explore the expected results of each protocol shown in Fig.~\ref{fig:S6}. The bit-purification circuit has the best outcomes. When using the phase error correction circuit, there is negligible improvement after purification, which is as expected, as the main errors are due to amplitude damping. In the simulations, double-selection purification introduced in Ref.~\cite{Fujii2009} has similar performance to bit-purification when the cable delay $t_d$ is small, but becomes worse when $t_d$ is large, resulting in larger (and therefore dominant) amplitude damping errors. Here we only perform purification that corrects damping errors, as this has the best simulated performance. More complex purification protocols \cite{Krastanov2019} could be explored if we had a better quantum memory (e.g. larger $T_1$ in the storage qubits), or if we had multiple communication channels, avoiding the need to store the first Bell pair.

\section{Discussion of Entanglement Protection with Dynamical Decoupling}
\begin{figure}
    \centering
    \includegraphics[width=0.8\textwidth]{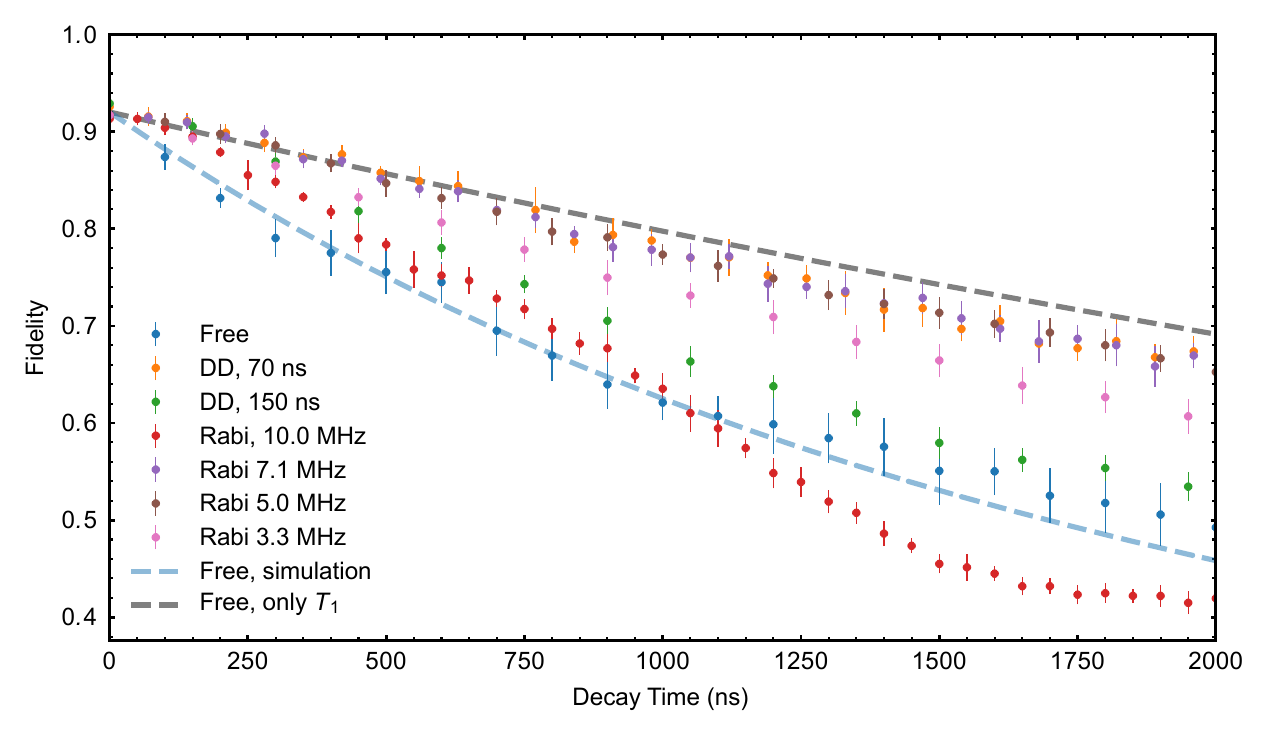}
    \caption{Bell state fidelity when varying the DD duration and Rabi drive strength.}\label{fig:S7}
\end{figure}

In Fig.~\ref{fig:S7} we plot the performance of the dynamical decoupling method for different DD durations and Rabi drive strengths. We change the interval between each $X$ gate in the DD sequence; the DD duration is the period for two complete $X$ gates. We find that for longer DD durations, the fidelity decays faster. We also vary the Rabi drive strength, where we find that for a larger drive strength $\Omega/2\pi = 10.0~\rm MHz$, the state fidelity decays much faster, which could be due to leakage to the higher excited states resulting in imperfect control of the qubit. For $\Omega/2\pi = 7.1~\rm MHz$, the performance is similar to the value $\Omega/2\pi = 5.0~\rm MHz$ used in the main text.

We also tried to combine RD/DD with entanglement purification to improve the performance of purification. However, we found in the experiment that to preserve the fidelity of the first Bell pair in $|Q_1^AQ_1^B\rangle$ during the generation of the second Bell pair, it is necessary to frequency-bias $Q_1^A$ away from $Q_2^A$ (see Fig. S4). This frequency bias requires that we apply a $z$ pulse to $Q_1^A$. We found experimentally that the DD/RD signals did not improve the Bell state fidelities when combined with the $z$ pulses. It could due to spurious noise introduced by the $z$ pulses, and/or imperfections in the pulse shapes. This meant we could not successfully combine DD/RD with purification.

